\documentclass[12pt,a4paper]{article}

\begin{document}
\title{Building Computer Network Attacks} 
\author{Ariel Futoransky, Luciano Notarfrancesco, \\
Gerardo Richarte and Carlos Sarraute
\\ \\ \textsc{CoreLabs, Core Security Technologies}}
\date{March 31st, 2003}
\maketitle

\begin{abstract}
In this work we start walking the path to a new perspective
for viewing cyberwarfare scenarios, by introducing conceptual
tools (a formal model) to evaluate the costs of an attack,
to describe the theater of operations, targets, missions, actions,
plans and assets involved in cyberwarfare attacks.
We also describe two applications of this model:
autonomous planning leading to automated penetration tests,
and attack simulations, allowing a system administrator to
evaluate the vulnerabilities of his network.

	
\end{abstract}

\section{Introduction}
In this work we set the basis of a framework for
modeling and building computer network attacks. 
The main purpose of this framework is to provide
a tool for automating the risk assessment process, 
in particular penetration tests, providing a further step in the direction
of a tool like Core Impact \cite{Co02}.

This work has also a theoretical value: 
``we understand what we can build.'' Our framework, 
considered as a functional model of the attacking process,
will provide the community with a deeper and more detailed model of the
attacks and intrusions of computer networks.

Finally, it can be used by a system administrator to simulate attacks against his network,
evaluate the vulnerabilities of the network and determine which countermeasures
will make it safe.

After reviewing related work, we describe in the second section the components of our model - 
probabilistic assets, quantified goals, agents and actions - and their relations.
In the third section we describe  the principal applications of this model:
automated planning of attacks and attack simulations.

\subsection*{Related work}

\subsubsection*{Description of security incidents}
In \cite{HL98}, Howard and Longstaff describe an incident taxonomy,
which was the result of a project to establish a
``common language'' in the field of computer security.
In \cite{LJ97}, Lindqvist and Jonsson also work on the classification 
of intrusions.
We try to use the high-level terms proposed by Howard and Longstaff,
in particular, the attributes: vulnerability, action, target, result and objective.

One common flaw of these classifications is that they
exclusively adopt the point of view of the system owners
or the intrusion detection sensors. But an attack is 
always composed of several steps, some of which may
be invisible to the sensors.
We will add to the attributes considered in \cite{HL98} and \cite{LJ97}, some
attributes which are relevant only to the attacker (or risk assessment team).
Thus a first result of our framework will be a widening of the concepts
used to describe security intrusions.

\subsubsection*{Attack models}
In \cite{Sc99} and \cite{Sc00}, Bruce Schneier proposes to describe attacks against a system
using ``attack trees'', where each node requires the execution of
children nodes, and the root node represents the goal of the attack.
There are two types of nodes: OR nodes and AND nodes.

In \cite{TLK01} the authors propose an attack specification language,
which extends the attack trees model. Each node has preconditions
(system environment properties which facilitate the execution of 
the attack node), subgoals (these are the children nodes), and
postconditions (changes in systems and environments). 
\cite{MEL01} is also based on the attack trees model. The model is extended
by attack patterns and attack profiles. These authors' objective is to
provide a means for documenting intrusions. Their models are purely
descriptive and doesn't allow us to construct or predict new attacks.

\subsubsection*{Attack graphs}
In \cite{SPG98} and \cite{JSW02} the authors propose to use attack graphs to determine the
security of networked systems. 
There are two main differences with our model. Firstly, the system they propose is 
an analysis system, which requires as input a great amount of information:
a database of common attacks broken in atomic steps, specific network configuration
and topology information, and an attacker profile. Our model is a system
for building attacks, starting with possibly zero information about the network,
and gathering this information as the attack takes place.
Secondly, the attack graph we construct for planning purposes differs from
the attack graph of \cite{SPG98} and \cite{JSW02}. In particular, it has a smaller size,
which allow us to effectively find a plan for real world instances of the problem.
Additionaly we introduce several new dimensions to the graph, like quantified goals, 
probabilistic assets and a complex cost function.

\section{Model architecture}

Our model of computer attacks is based on the concepts of
assets, goals, agents and actions.
The actions are conceptual units: they are the building blocks of the attacks.
In our description, an attack involves a group of agents, 
executing sequences of  actions, 
obtaining assets (which can be information about the network 
or actual modifications of the real network) in order to reach a set of goals.
A functional prototype of this framework was written in 
Squeak Smalltalk, and we occasionally use the Smalltalk vocabulary
by speaking of classes, instances, variables and methods.

\subsection{Assets}

An asset can represent anything that an attacker may need to
obtain during the course of an attack. More precisely, it 
represents the knowledge that an agent has of a real object 
or property of the network. Examples of assets are: \\
{\tt 
\indent * AgentAsset (agent, capabilities, host) \\
\indent * BannerAsset (banner, host, port) \\
\indent * OperatingSystemAsset (os, host) \\
\indent * IPConnectivityAsset (source, target) \\
\indent * TCPConnectivityAsset (source, target, port) }

\noindent
An AgentAsset represents an {\tt agent} with a collection of {\tt capabilities}
running on a {\tt host}. A BannerAsset represents the {\tt banner} that an agent
obtains when trying to connect to a certain {\tt port} on a {\tt host}. 
An OperatingSystemAsset represents the knowledge that an agent
has about the operating system of a {\tt host}.
A TCPConnectivityAsset represents the fact that an agent is able
to establish a TCP connection between a {\tt source} host and a certain
{\tt port} of a {\tt target} host.

There is an implicit question associated with each asset.
For example, an OperatingSystemAsset with {\tt os=nil} and
{\tt host=192.168.13.1} is associated with the question
``What is the operating system of the machine whose IP is 192.168.13.1 ?".
This gives us a natural relation between assets of the same class : 
an asset $a1$ completes another asset $a2$ if $a1$ has some extra 
information which answers the question implicitly associated with $a2$.
In our example, if two OperatingSystemAssets $a1$ and $a2$ have
the same {\tt host=192.168.13.1}, $a2$ has {\tt os=nil} and
$a1$ host {\tt os=linux}, $a1$ completes $a2$.

The assets we consider are probabilistic. This allow us to represent
properties which we guess are true with a certain probability
or negative properties (which we know to be false). For example,
an action which determines the os of a host using banners 
(OSDetectByBannerGrabber) may give as result an 
OperatingSystemAsset {\tt os=linux} with {\tt probability=0.8} and
a second one with {\tt os=openbsd} and {\tt probability=0.2}. 
Another example, an ApplicationAsset {\tt host=192.168.13.1}
and {\tt application=\#Apache} with {\tt probability=0} means that 
our agent has determined that this host is not running Apache.

We associate to each asset a certain level of trust in the information it represents.
When an action returns an asset, we may trust this information, 
but then trust diminishes with time. This decrease is not linear - 
how to calculate it is an interesting open question.

\subsubsection*{The environment knowledge}

The environment knowledge is a collection of information
about the computer network being attacked or hosting an agent.
Naturally, this information is represented by assets.
By abuse of language, we may speak of the {\em environment} instead
of the environment knowledge.
In the beginning, the environment contains only an AgentAsset : 
the {\tt localAgent} which will initiate the attack.

The environment will play an important role during the planning phase
and during the execution phase of an attack, since it 
continuously feedbacks the behavior of the agent.
Note also that each agent has its own {\em environment knowledge}
and that exchanging assets will be an important part of the
communications between agents.

\subsection{Goals}

A goal represents a question or a request of the type:
``complete this asset" (every goal has an associated asset).
A goal is quantified (has an ordered collection of quantifiers).
A goal also knows all the actions which may complete his asset.
This list is filled out during the attack graph construction phase.

\subsubsection*{Quantifiers}
We consider three types of quantifiers: {\em Any},
{\em All} and {\em AllPossible}. An example will clarify
their meaning: consider that PortAsset has attributes 
({\tt host, port, status}).
The following goals will mean: 

\noindent
{\tt asset = PortAsset (host=192.168.13.1, status=\#open), \\
quantifiers = (Any \#port from:1 to:1024)}: \\
find an open port in host 192.168.13.1 between ports 1 and 1024.
To fulfill this goal, an action like PortScan will begin examining
the ports of host 192.168.13.1 until it finds an open port (completes
the PortAsset and returns a success signal) - or reaches
port 1024 (and returns a failure signal). 

\noindent
{\tt asset = PortAsset (host=192.168.13.1, status=\#open), \\
quantifiers = (All \#port from:\#(21,22,23,80))}: \\
find whether all the ports \#(21,22,23,80) are open in host 192.168.13.1.
This time, PortScan will examine the four mentioned ports and return
success only if the four of them are open (and in that case
completes four PortAssets).

\noindent
{\tt asset = AgentAsset (capabilities=\#(TCP,UDP,FileSystem)), \\
quantifiers = (AllPossible \#host from:192.168.1.0/24)}:\\
install agents in all the hosts that you can in netblock 192.168.1.0/24.
An action able to fulfill this goal will be a subclass of Exploit
(for example ApacheChunkedEncodingExploit). To fulfill this goal, the Exploit 
action will try to exploit a vulnerability in all the machines it reaches
in that netblock. It returns an AgentAsset for each compromised host,
and returns success if at least one machine is compromised.

In general, the quantifiers are an ordered collection. Of course, the order
is important: for example, a goal with {\tt asset = PortAsset (status=\#open)}
and quantifiers {\tt (All \#host from:192.168.1.0/24), (Any \#port from:1 to:1024)}
will mean: for every host in this netblock, find an open port. 
Whereas with quantifiers {\tt (Any \#port from:1 to:1024), All \#host from: 192.168.1.0/24)}
it will mean: find a port which is open in all the machines in this netblock.

\subsection{Attackers and agents}

We can think of the actions as being the verbs in a sequence of 
sentences describing the attack. The agents will then be the subject
of those verbs.
Of course, an attack is always initiated by human attackers,
but during the course of the attack, actions will typically be executed 
by autonomous agents.

\subsubsection*{Human attackers}

There are different types of attackers. 
They can be classified grosso modo as :
script kiddies, who attack more or less randomly 
using standard tools downloaded from the Internet; 
hackers, who attack computers for challenge, status or 
research;
security auditors (pen testers), who evaluate the security
of a network;
government agencies, who possess the highest skill level
and resources to perform an attack.

The way we model these different types of attackers is
through the attack parameters: stealthiness, non traceability,
expected running time, expected success; and a skill level
given by the collection of actions available to the attacker.
A script kiddie will not worry about stealthiness or non traceability.
His attacks will have low expected success and require low
skill level. On the other hand, a government agency will use
maximal stealthiness, non traceability and skill level, with
a high expected success. A security auditor will not worry
about non traceability but may require stealthiness to carry on
the penetration test.

There is a number of actions who will require a human agent
to execute them, for example social engineering.\footnote
{Notice that social engineering could also be performed by an
autonomous agent. An example of this would be a virus who relies
on a suggestive title to be opened by the receiver.}
But it is important to include them in our model - 
for the sake of completeness, but also because they are
necessary if we want to do simulations using our framework.

\subsubsection*{Software agents}

In general the execution of an action will require the execution
of machine code and therefore involves a software agent $A$
executing this code. The command of executing this action
might come from the agent itself, from another
software agent or from a human attacker:
we will not distinguish between these cases and say 
that the action was executed by the agent $A$.
A software agent can take several forms: script, toolkit or other
kinds of programs. Let us point out the {\em autonomous agents}
who are able to take decisions and continue the attack
without human intervention.

\subsubsection*{Communications between agents}

This framework supports the interactions between agents, 
which collaborate to accomplish the objective.
The agents establish communication channels between them
to exchange knowledge, gained information and
missions to be executed.
For example, each agent has a collection of actions. 
Agents can learn new actions and  exchange actions between them through the 
communication channels.

The communications between human attackers can take place
through unlimited types of channels (direct interaction, telephone,
email, chat, irc, written mail). We will not enter into details here.
Examples of communication channels between software agents
are POP3 and SMTP, HTTP/HTTPS, DNS, and covert channels like Loki.

\subsubsection*{Agent mission}

We contemplate different types of organizations between the agents.
One scenario is given by a ``root agent'' who plans the attack and
then gives the other agents orders (of executing actions), eventually
creating new agents if necessary, and asks the agents for 
feedback about action results in order to decide further steps.

Another scenario is when the root agent delegates responsibilities
to the other agents, giving them higher level {\em missions}.
To fulfill the mission, the agent will have to do his own planning
and communicate with other agents. This scenario is likely
to arise when stealthiness is a priority: communications are very
expensive and it becomes necessary to rely on the agents to
execute their missions without giving feedback (or the smallest
amount of feedback, or delayed feedback because of 
intermittent communication channels).

\subsection{Attack actions}

These are the basic steps which form an attack.
Examples of actions are:
ApacheChunkedEncodingExploit, WuFTPglobbingExploit
(subclasses of Exploit), BannerGraber, OSDetectByBanner,
OSFingerprint, NetworkDiscovery, IPConnect, TCPConnect.
In this section we review the principal attributes of an action.

\subsubsection*{Action goal}

An action has a goal (naturally an instance of the {\em Goal} class previously 
described) and when executed successfully the action completes the asset
associated with its goal.

Usually,\footnote{For example see \cite{TLK01}}
an action is directed against a target,
where the target is a computer or a network.
But as we have seen in the {\em Goals} section,
there are different types of goals like gathering information
or establishing connectivity between two agents or hosts,
where the notion of target is not so clear. Thus the concept
of goal is more general and allow us to speak about intermediate
steps of an attack.

It is also common to speak about the result of an action
(for example to increase access, obtain information, corrupt information, 
gain use of resources, denial of service), focussing on non authorized results.
Our concept of goal also includes this idea.
Note that when an action completes the goal asset,
we are taking into account only the expected result of the action.
Undesired results and other side effects fall into the category of noise.

Also, thinking about goals opens the door to a whole new set
of tools and ideas, giving a new dimension to the available
actions for planning and developing attacks. An important
example is the idea of Persons as targets.
Currently the weakest link in the security of an organization,
for different reasons, is the workstation \cite{AL03}. When
attacking servers, the adversary is the system administrator
or even the security professional in charge of it, but when
attacking workstations, the adversary is the end user, who probably
doesn't fully understand the implications of his actions, raising
the possibilities of finding a vulnerability or configuration problem,
and it's most likely to stay undiscovered after the intrusion.

Before assigning a goal to an action, the action remains abstract
(described by the methods of its class). Given a goal, we can
instantiate the action, but because of the quantifiers this instantiation
is only gradual. During the planning phase, actions are only
partially instantiated and use the common information of the goal asset
(for example if the goal is to find AllPossible open ports in a netblock,
we only use the fact that we have to find open ports).
The {\em initializeRequirements} method is called, which creates the list
of requirements or subgoals of the action, needed to carry on the 
construction of the attack graph.
During the execution phase, when quantifiers are recursively iterated,
the action is fully instantiated and receives a concrete asset to fill out (for example, to
determine if a specific port of a specific host is open). 
The {\em setupRequirements} method is called, which communicates
this information to the subgoals of the action.

\subsubsection*{Action requirements}
The {\em requirements} are instances of the Goal class,
and will be the goals of other attack actions, which must have been successfully
executed before the considered action can be run. 
The requirements are the equivalent of children nodes in \cite{Sc00}
and subgoals in \cite{TLK01} and \cite{MEL01}.
An abstract action must know what kind of assets it may satisfy
and which goals it requires before running. These relations will be
used to construct the attack graph.

\subsubsection*{The {\em run} method or the action itself}
During the execution phase, when the action goal is fully
instantiated, the {\em run} method is called. This is of course
the most important method of an action, which contains
the concrete code executed by the agent.
Before executing its own code, the action will call
the {\em runAction} method of its requirements.

One interesting point to mention here, is that the action
can be executed in a real network or in a simulated network
(with simulated hosts and network topology). The difference
between working in those two settings will be reflected
only in the {\em run} method of the actions. This makes our 
framework easy to adapt for both real and simulated attacks.

\subsubsection*{Environment knowledge}
The action makes use of the knowledge that his owner agent has 
of the environment.  This environment knowledge is a collection of information
about the computer network being attacked or hosting the agent.
When the {\em run} method is called, the first thing that the action
does is to examine the environment, looking for an asset which
completes the goal asset. If this is the case, the information of 
the existing asset is used to fulfill the goal, and the action returns
a success signal, resulting in zero cost (in terms of time, noise, success 
probability and stealthiness).

Note that two interesting graphs can be extracted from the environment knowledge:
the network topology graph and 
the agent distribution graph, whose nodes are the agents involved in the attack
and whose edges are the communication channels between agents.

\subsubsection*{Environment conditions}
The environment conditions refer
to system configuration or environment properties which may be
necessary or may facilitate the execution of the action.
We distinguish the environment conditions from the requirements,
because the requirements express relations between actions
(which must be taken into account when planning a sequence of actions)
whereas the environment conditions refer to the ``state of the world"
(as far as the agent is aware of it) before the execution of the module,
and do not imply previous actions. 
For example, an exploit of a buffer overflow that runs only on specific versions
of an operating system, will have as requirement: ``obtain information
about operating system version'' and as environment condition
``OS=RedHat Linux; version between 6.1 and 6.9''. These conditions are
not necessary, as the action can be run anyway, but will dramatically
increase its probability of success.\footnote
{Note that the system administrator
could have changed the version number of the OS, instead of upgrading it,
so the exploit could run although the environment conditions indicate
it will fail. In this situation, executing the exploit or not will depend on other
global parameters: noise level, expected success, execution time.}

\subsubsection*{Noise produced and stealthiness}
The execution of the action will produce {\em noise}. This noise
can be network traffic, log lines in IDS, etc. 
Given a list as complete as possible of network sensors,
we have to quantify the noise produced respect to each of this sensors.
The knowledge of the network configuration and which sensors
are likely to be active, will allow us to calculate a global estimate of 
the noise produced by the action. Refining this estimation is an
interesting open question.

With respect to every network sensor, the noise produced can be
classified into three categories: unremovable noise, noise that can
be cleaned in case the action is successful (or another subsequent
action is successful), noise that can be cleaned even in
case of failure. So we can also estimate the noise remaining after
cleanup. Of course, the {\em stealthiness} of an action will refer
to the low level of noise produced.

\subsubsection*{Exploited vulnerability}
The module, if aiming to obtain an unauthorized result, will exploit
a vulnerability. The information about the {\em exploited vulnerability} is
not needed by the attacking agent, but is useful for the classification and
analysis of detected intrusions.  This vulnerability can be:\\
\indent * software vulnerability: a design flaw or an implementation flaw 
(buffer overflow, format string, race condition)\\
\indent * network configuration vulnerability\\
\indent * trust relationship: this refers to higher level, non autonomous
attack modules: hacking a software provider, getting an insider in a software
provider, inserting backdoors in an open-source project.

\subsubsection*{Running time and probability of success}
The {\em expected running time} and {\em probability of success}
depend on the nature of the action, but also
on the environment conditions, so their values must be updated
every time the agent receives new information about the environment.
These values are necessary to take decisions and choose a path
in the graph of possible actions. Because of the uncertainties
on the execution environment, we consider three values for the
running time: minimum, average and maximum running time.
Together with the stealthiness and zero-dayness,
these values constitute the cost of the action
and are used to evaluate sequences of actions.

\subsection{Building an attack}

An attack involves a group of agents, executing series of actions
in order to fulfill a goal (or a set of goals).
Notice that this goal can change during the course of the attack.

The {\em target} is the logical or physical entity which is the blank
of the attack. Usually, the target is a computer or a 
computer network or some information hosted in a computer.
The target can also change during the course of the attack.
It is also possible that an attack has no specific target at all
(for example, a script kiddie running a specific exploit
against all the computers he reaches, until it succeeds).

The complete graph of all combinations of actions determines which goals
we (as attackers) can reach. Considering the complete graph
of possible actions, to build an attack will consist in 
finding a path to reach the objective (this implies in particular
to find a path through the physical networks to reach the target).

\subsubsection*{Attack parameters}
In fact, we will try to find the best path to reach the objective,
and to evaluate this we must take into account the attack
parameters: non traceability, tolerated noise, expected success,
execution time, zero-dayness.
These parameters have initial values for 
the whole attack, but they can vary from agent to agent,
for example an agent might create other agents with a different profile.

{\em Non traceability}  refers to the ability to dissimulate the 
origin of the attack. We could also call it ``deniability''.
To achieve non traceability, special modules must be
executed, who will insert intermediate agents
(we call them ``pivots'' or stepping stones) between
the originating agent and the agents executing the
objective or partial objectives.

{\em Tolerated noise} is the level of noise that we allow
our agents to make. It can vary from agent to agent,
for example an agent executing a DNS spoofing module
would benefit from other agents simultaneously
executing a DNS flooding (and generating a high level
of noise).

{\em Expected success} determines the priority which
will be given to successfully executing the actions, 
over the other parameters. If set to the maximal value,
the agent will try to execute the projected action,
even though the noise generated might be higher
than the tolerated noise level.

{\em Execution time}$\,$: each agent will be given a
limit of time to execute each action. This is necessary
to plan the attack, as it usually consists of a series of dependent
steps.

{\em Zero-dayness}$\,$: specifies whether the agent is allowed to use
zero-day exploits (a valuable resource that should be used only
for critical missions).

\subsubsection*{Evaluating paths}
A path is a sequence of actions (in a specific order and without branchings).
To be able to choose between different paths,
we have to evaluate the paths in terms of the attack parameters:
the probability of success, the noise probably produced,
the running time and the traceability.

For the probability of success, we consider
the success of each action as independent of the previous ones,
and that all the actions in the paths are necessary,
so the probability of success of the path is the product of
the probabilities of success of each action.

For the running time of the path, we consider the three
time estimations (minimum, average, maximum) 
that we have for each module and sum them to obtain
the path's minimal, average and maximal running time.

The stealthiness of the path, that we define as the 
probability of not being discovered, diminishes with
each executed action. As with the probability of success,
we consider them independent and compute the stealthiness
of the path as the product of the stealthiness of the actions.

The traceability is harder to estimate. It depends 
basically in the number of hops (or pivots or stepping stones)
introduced, and this is how we compute it, although each can contribute in a 
different amount to the global ``non traceability'' of the path of actions.

\section{Applications}

\subsection{Planning}

\subsubsection*{The complete graph of possible actions}

Our first approach to the planning problem is to contemplate
the complete graph of possible actions.
In this graph, each node represents a {\em``state of the world''},
which is the combined knowledge that the agents have
of the environment (network topology, operating systems,
running services, agents distribution, etc).
Each edge represents an action (the execution of a module
by an agent). This action will result in modifying the
{\em state of the world} (which can be changes in the real world,
or changes in the knowledge the agents have about the world).

This model allows us the express complex relationships between
actions, and always takes into account the order in which they are 
executed. In this sense, it is close to reality. 
But autonomous planning is untractable: it is only possible
for very small instances of the problem.

\subsubsection*{Probabilistic planning}

Another approach is to describe the model as a STRIPS-like planning
domain where the actions have probabilistic outcomes.
This is the setting of Markov Decision Processes (MDP).
To do this, we represent our modules as probabilistic operators
and the results and noise produced by a module as 
add-effects and delete-effects of the operator.
These effects depend on a probability distribution.
The requirements and environment conditions of
the module are translated as preconditions of the operator.

One difficulty is that the planning operators
do not create or destroy objects, so we must begin planning
when we already know the topology of the computer network
being attacked. The description of the network will then be
the initial conditions. 

Having described the planning of the attack as a Markov Decision Process,
we can use existing tools for solving such MDPs. For example,
the probabilistic planner {\em PGraphPlan} presented in \cite{BL98},
based on the deterministic planner {\em GraphPlan} \cite{BF95}.
Although this planner is quite efficient,
it only solves the planning problem when the plan requires
a small amount of steps.

\subsubsection*{Our approach}
We construct the attack graph by alternating layers of goals and
layers of actions: we connect each goal in a layer to
the actions which may satisfy it, and connect each action
in a layer to the subgoals that it requires.
Note that we don't need to build explicitly this graph,
in fact we prefer to leave it implicitly defined by the relations
between actions and requirements.

As we have already stated, we consider two different phases:
during the attack graph construction, the quantifiers are not
expanded, and the collection of elements referred by the 
quantifiers are treated as a single element (in this sense, 
we say that the elements are homogeneous).
This use of quantifiers is a great help in avoiding the combinatorial
explosion of states, a fundamental problem for the planning systems.
This is also a way to take into account the uncertainty that we have
respect to the real world: a collection of elements (hosts, ports, applications)
that we cannot distinguish because of lack of information, are in the
beginning treated as the same element, thus facilitating the planning,
and are distinguished only after enough information is gathered.

During the execution phase, when the {\em run} method
is called, the action recursively iterates over the quantifiers, thus
delaying the expansion until it is absolutely necessary.

Another thing that we delay is the computation of the costs of the actions.
In effect, the cost is greatly influenced by the environment knowledge,
and this knowledge is modified by every actions which
gets executed. It is only when the {\em runAction} method of a goal is called,
that the goal computes the costs of the actions which may satisfy it, and
chooses the action with the lower cost. This way, the same attack graph
will result in different executions if we run it several times without modifications:
for example after running an ApacheChunkedEncodingExploit in a first execution,
the exploit may add a negative asset (with probability zero) to the environment
stating that the target host is not running Apache. In the following execution,
the probability of success of this exploit will be almost zero and another exploit
will be executed.

An interesting issue is raised by the fact that an agent may create 
another agent. For example given the goal of establishing a TCPConnectivity,
a TCPConnect action requires to install an agent in an intermediate host
in order to circumvent a firewall. To install this agent, an exploit requires
a TCPConnectivity, which in turn requires to install an agent in an intermediate host,
etc... we are faced with combinatorial explosion and infinite loops.
To solve this issue we have used an idea which resulted to be a common technique
in Hierarchical Task Network (HTN) planning. 
As stated in \cite{SFJ00}, virtually all planning systems that have
been developed for practical applications make use of HTN planning techniques.
As long as the actions do not require to create new agents, we use classical planning
techniques. But when an action such as TCPConnectCreatingHops requires installation
of new agents, we consider it as a high-level task. TCPConnectCreatingHops will
find the best path between our agents and a target host, eventually winning
new agents in intermediate machines. If we can compute the cost of the connectivity
between two hosts and the cost of ``winning an agent in host {\em target} from 
host {\em source}", we can use a standard algorithm like the algorithm of Dijkstra
to find the best path in the network topology graph. To evaluate the cost of
winning an agent, we construct a subplan using an environment containing
only the {\em source} and {\em target} hosts. The cost of this subplan is computed
by appending to this hypothetic environment the real environment together with
an imaginary agent in host {\em source} (this is the way we can work out hypotheses).
This subplan will also be used during the execution phase, each time that an 
agent must be won in the path found by the Dijkstra algorithm.

\subsection{Simulations and analysis of network security}

As we have mentioned in the {\em Actions} section, our framework
can be used to build attacks against a simulated network.
Of course, the quality of this simulation will depend on how accurately
we simulate the machines. Using VMwares we obtain a slow and accurate
simulation, for faster simulations a tradeoff must be made between accuracy
and speed.

The system administrator can simulate different types of attackers
by using different attack parameters and different collections of available actions,
and evaluate the response of his network to these attackers.
For example, he can start with an attacker with a minimal portfolio of actions,
and add gradually actions to the arsenal of his simulated attacker
until there is a successful attack which goes undetected by the IDS.
This will give him an indication of which attack actions he should
defend his network from.

Also consider that the system administrator has a set of measures
which make certain attack actions less effective (in our framework, a measure
may reduce the probability of success of an attack action, or increase
the noise it produces, for example by adding a new IDS).
He can then use the simulation to see if his system becomes safe
after all the measures are deployed, or to find a minimal set
of measures which make his system safe.

As opposed to VMwares, rudimentary simulations of machines allow
us to simulate a huge amount of machines. This can be used to
investigate the dissemination of worms (considered as autonomous
agents with a minimalist set of actions) in large networks.  Future work
needs to be done in this direction.

\section{Conclusions and future work}

In this paper we start walking the path to a new perspective for 
viewing cyberwarfare scenarios, by introducing different conceptual tools
(a formal model) to evaluate the costs of an attack, to describe the theater
of operations, targets, missions, actions, plans and assets involved in 
cybernetic attacks.

Thinking about the way to express goals, led us to introduce three different
quantifiers (Any, All, AllPossible) and expand the notion of target of an action.
The quantifiers give us a compact representation of extensive collections
of assets, which greatly reduces the combinatorial growth of the attack graph
during the planning phase.

Another important contribution concerns the costs of the actions.
We show that the cost is given by a tuple a values: not only the
probability of success, but also the stealthiness (which depends
on the noise produced), time consumed, non traceability and
zero-dayness. The noise produced is particularly relevant, and
we haven't seen it in other models. These dimensions, considered
as attack parameters, also allow us to model different types of attackers.

The most important application of our model is automated planning.
Integrated in a tool like Core Impact, it leads the way to automated
penetration testing. Used against simulated networks, it is a 
tool for evaluating the vulnerabilities of a network.

In our approach to the planning issue, we chose not to rely entirely
in one classical representation (like STRIPS or MDP representations,
or formal models used in model checking) and then use standard tools
for this representation, because these only work for small instances
of the problem. By combining classical planning techniques with 
HTN techniques, and making effective use of the quantifiers,
we were able to handle bigger instances of the problem
in our Squeak implementation (not particularly optimized!)

Finally working on this model has opened a lot of questions and
directions for future work:
how to estimate the probability of success and noise produced by
an action, how to modify these values after an execution,
how to combine the different dimensions of the cost in order
to obtain a total order between costs,
how to choose the agents who will execute the actions,
when to create a new agent on a specific host,
how to decide the profiles (or personalities) of the agents,
the use of planning techniques, 
and the applications of the simulation scenario.
It also led us to review current penetration testing practices
and opened new dimensions for planning attacks and creating new tools.


\end{document}